\newcommand*{\aZ}{}

\ifdefined\aZ

\documentclass[11pt]{article}
\usepackage{graphicx}
\usepackage[lmargin=1.0in, rmargin=1.0in,tmargin=2cm,bmargin=2.50cm]{geometry}
\usepackage[T1]{fontenc}
\usepackage[utf8]{inputenc}
\usepackage{authblk}
\usepackage{xcolor} 
\usepackage{booktabs}

\else

\documentclass{tEWA2e}

\usepackage{xcolor}

\fi  

\def\Xint#1{\mathchoice
{\XXint\displaystyle\textstyle{#1}}%
{\XXint\textstyle\scriptstyle{#1}}%
{\XXint\scriptstyle\scriptscriptstyle{#1}}%
{\XXint\scriptscriptstyle\scriptscriptstyle{#1}}%
\!\int}
\def\XXint#1#2#3{{\setbox0=\hbox{$#1{#2#3}{\int}$}
\vcenter{\hbox{$#2#3$}}\kern-.5\wd0}}

\def\dashint{\Xint-}
\makeatletter
\def\maketag@@@#1{\hbox{\m@th\normalfont\normalsize#1}}
\makeatother

\begin{document}

\ifdefined\aZ

\title{\bf{Comparison of iterative solvers for electromagnetic analysis of
plasmonic nanostructures using multiple surface integral equation formulations}}
\author[1]{Hip\'olito G\'omez-Sousa}
\author[1]{\'Oscar Rubi\~nos-L\'opez}
\author[2]{Jos\'e \'Angel Mart\'inez-Lorenzo}
\affil[1]{Department of Signal Theory and Communications, University of Vigo,\protect\\EI de Telecomunicaci\'{o}n, ES~36310 Vigo, Spain.\protect\\{\color{blue}{\{hgomez, oscar\}@com.uvigo.es}}}
\affil[2]{Northeastern University, 360 Huntington Ave.,\protect\\Suite 302 Stearns Center, Boston, MA~02115, USA.\protect\\{\color{blue}jmartinez@coe.neu.edu}}
\renewcommand\Authands{ and }
\date{}
\maketitle

\else


\title{Comparison of iterative solvers for electromagnetic analysis of
plasmonic nanostructures using multiple surface integral equation formulations}

\author{H. G\'omez-Sousa$^{\rm a}$,
\'O. Rubi\~nos-L\'opez$^{\rm a}$$^{\ast}$\thanks{$^\ast$Corresponding author. Email: oscar@com.uvigo.es
\vspace{6pt}}
and J. \'A. Mart\'inez-Lorenzo$^{\rm b}$$^{\ast}$\thanks{$^\ast$Corresponding author. Email: jmartinez@coe.neu.edu
\vspace{6pt}}
\\\vspace{6pt}
$^{a}${\em{Dept. of Signal Theory and Communications, University of Vigo, EI de Telecomunicaci\'{o}n, ES~36310 Vigo, Spain}};
$^{b}${\em{Northeastern University, 360 Huntington Ave., Suite 302 Stearns Center, Boston, MA~02115, USA}}
\\\received{v1.1 released September 2015} }


\maketitle

\fi

\begin{abstract}
The electromagnetic behavior of plasmonic structures can be
predicted after discretizing and solving a linear system of equations, derived
from a continuous surface integral equation (SIE) and the appropriate boundary
conditions, using a method of moments (MoM) methodology. In realistic large-scale
optical problems, a direct inversion of the SIE--MoM matrix cannot be performed
due to its large size, and an iterative solver must be used instead. This paper
investigates the performance of four iterative solvers (GMRES, TFQMR, CGS, and
BICGSTAB) for five different SIE--MoM formulations (PMCHWT, JMCFIE, CTF, CNF, and
MNMF). Moreover, under this plasmonic context, a set of suggested guidelines are
provided to choose a suitable SIE formulation and iterative solver depending on
the desired simulation error and available runtime resources.

\ifdefined\aZ

\else


\fi

\end{abstract}

\section{Introduction}

During the last years, we have witnessed an increasing interest in the optical
properties of metallic nanoparticles. Research in the area of plasmonic
nanoelectronics explores the behavior of electromagnetic fields which are
confined over dimensions smaller than the wavelength. This field confinement is
induced by interactions between electromagnetic waves and conduction electrons at
the interfaces of metallic nanostructures. These plasmonic interactions can be
used to manipulate light beyond the classical diffraction limit, giving rise to
an emerging wide range of interesting novel applications in sensing and
waveguiding, amongst other applications \cite{1,2}. Additionally, at optical
frequencies, metallic nanostructures can be characterized by localized
surface--plasmon resonances (LSPRs) that support a strong enhancement in the
directivity of spontaneous emission of light by single fluorescent molecules or
other point-like emitters of light. This enhancement in the emission of light is
a fundamental aspect that has recently originated an intensive promising research
work on the design and the experimental construction and testing of nanoantennas
[3--7]. As a consequence of this novel research area pioneered in recent years,
numerical techniques for providing accurate simulation and analysis of problems
involving plasmonic nanostructures are indispensably required to exploit the
growing range of applications relying on optical plasmonic properties.

The behavior of nanoparticles at optical frequencies can be well modeled by
classical electrodynamics [1]. Nevertheless, in electromagnetic optics, the penetration of fields
into metals must be considered and therefore SIE (surface integral equation)
formulations for penetrable scatterers are a suitable choice for simulating the
optical responses of isolated plasmonic nanoparticles or composite structures
consisting of multiple nanoparticles. Surface integral equation techniques based
on the method of moments (MoM) have demonstrated to provide very accurate
simulation results in many different problems involving real plasmonic objects
[8--11]. In spite of being a classical approach, the SIE--MoM method delivers
accurate predictive results for particle and surface feature sizes down to
$\sim1$~nm, a distance below which quantum non-local phenomena become
non-negligible [11]. Although not yet widely employed in optics, the SIE--MoM
approach brings important advantages over volumetric approaches such as the
discrete-dipole approximation (DDA) [12], the finite difference in time domain
(FDTD) method [13], and the frequency-domain finite-element (FEM) methods [14].
Such advantages include the fact that the SIE--MoM approach requires
discretizations of the material boundary surfaces only, thus generally reducing
the number of unknowns with respect to integral formulations based on volumetric
mesh modeling. Additionally, absorbing boundary conditions and surrounding empty
space need not to be parametrized, resulting in a significantly easier mesh
modeling. Finally, the SIE--MoM methods are less prone to be affected by
instabilities due to abrupt and spatially rapid variations of the permittivity
[11], as commonly occurs in the case of plasmonic problems.

There are many known factors that determine the final accuracy and the total
runtime when simulating with the SIE--MoM approach. However, two of them are of
special interest when solving this type of problems: the SIE--MoM formulation, and
the iterative numerical solver. Previous studies have concentrated on assessing
which SIE formulations are more suitable for different types of electromagnetic
scattering and radiation problems in terms of both runtime and accuracy [15--20],
including problems involving plasmonic structures [20]. Nevertheless, there is a
general lack of publications that focus explicitly on the role of the iterative
solver. Addressing this lack of results would certainly be relevant as the selected iterative
solver has a strong impact on the total simulation runtime and memory
consumption. This impact is particularly important when dealing with real-world
plasmonic problems, because they usually require performing massive batches of
large-scale simulations for different varying parameters such as optical
wavelength, illumination beam, and relative permittivity.

A comparative study of the performance of some well-known SIE formulations was
first released in [15] for non-plasmonic conducting and dielectric objects. Later
on [16], this was also done for perfect electric conductor (PEC) bodies. A
similar study for dielectric objects, but using acceleration algorithms for MoM,
was carried out in [17]. A comparative study applying the SIE--MoM approach to
left-handed metamaterials (LHM's) was reported in [18]; and, later on, this was
done using acceleration algorithms [19]. The foremost published comparison of
five widespread SIE formulations in the interesting context of plasmonic media
was accomplished in [20]. All the above-mentioned publications also include
iterative-performance studies. However, these studies are only focused on a
single iterative solver. The iterative-performance study carried out in [15] was
realized using only as iterative solver the ordinary (without restarts) version
of the generalized minimal residual method (GMRES) [21]. A restarted GMRES with
fixed restart parameter was chosen in \cite{16, 18, 20}, whereas another iterative
solver called BICGSTAB [22] was used for all the analyses published in \cite{17, 19}.

As detailed in 1.2.5 of [23], the algorithmic performance and the memory
complexities of each iterative solver may vary significantly depending on the
considered type of engineering problem. As a consequence, in order to optimize
the performance and memory usage for a given class of problems, there are no general
selection rules for the solver and an investigation should be required to compare
iterative solutions provided by various algorithms. This kind of important study
was left out in all the iterative-performance analyses reported in the aforesaid
publications [15--20], as these papers perform comparative analyses by only
varying the SIE formulation for a fixed iterative solver. In the present work, we
extend to multiple iterative solvers some comparative multiple-SIE single-solver
results published in [20] for plasmonic scatterers. Furthermore, a memory
complexity analysis not included in the previously mentioned reference is
presented in this paper. Moreover, the nanoscatterers used for obtaining the
results in [20] consist only of spheres simulated at a single operating
frequency. The selection of spheres as targets is well justified by the
availability of the Mie's series analytical reference results [24]; however, in
the present paper we also analyze the iterative performance results when dealing
with more elaborate geometries, such as a real plasmonic nanoantenna. In addition
to the study of the iterative performance, we also check, following a procedure
similar to that in [20], the accuracy of each SIE formulation when the MoM linear
system is solved by different iterative techniques.

The SIE formulations considered for the iterative performance comparison in this
paper include tangential equations only (combined tangential formulation, known
as CTF, and Poggio-Miller-Chang-Harrington-Wu-Tsai formulation, named by the
acronym PMCHWT), normal equations only (combined normal formulation, CNF, and
modified normal M\"{u}ller formulation, MNMF) and both normal and tangential
equations (electric and magnetic current combined-field integral equation,
JMCFIE). A detailed description of these five formulations can be found in
\cite{20,23}. The four studied iterative methods, based on Krylov subspaces, used for
solving linear systems of equations are the following: GMRES (generalized minimum
residual method), TFQMR (transpose free quasi-minimal residual method), CGS
(conjugate gradients squared method) and BICGSTAB (bi-conjugate gradients
stabilized method). Thorough explanations for all these solvers can be found in
[22]. In the particular case of GMRES, in addition to the iterative performance,
we analyze the influence of the restart parameter on the memory consumption as
well.

The rest of this paper is organized as follows. The SIE formulations considered
in the study are described in Section 2. Next, in Section 3, a concise overview
on Krylov iterative methods applied to SIE formulations is given. We present in
Section 4 a comparative study of numerical results for the considered SIE
formulations. First, in Subsection 4.1, representative sets of near and far field
outcomes obtained employing each SIE formulation are compared with the analytical
results provided by the Mie's series, in order to determine the level of accuracy
for each formulation. Then, in Subsection 4.2, the iterative performance of each
linear solver is assessed by measuring the runtime to solve the MoM system and
the total number of required matrix-vector multiplications for different
plasmonic geometries. Finally, Section 5 concludes the paper with a summary.

\section{Surface integral equation formulations}

In the analysis of the scattered field produced by an impinging electromagnetic
wave on a penetrable object, both tangential and normal boundary conditions are
typically imposed for the electric and magnetic fields at the interface of the
object. These boundary conditions establish the normal electric field integral
equation (N--EFIE), the normal magnetic field integral equation (N--MFIE), the
tangential electric field integral equation (T--EFIE) and the tangential magnetic
field integral equation (T--MFIE). The following linear combinations of the
aforementioned formulations are known to provide stable sets of SIE formulations
\cite{20, 23}:
\begin{equation}
\begin{array}{c}
\frac{{{a_1}}}{{{\eta _1}}}\left( {{\rm{T}} - {\rm{EFI}}{{\rm{E}}_1}} \right) +
\frac{{{a_2}}}{{{\eta _2}}}\left( {{\rm{T}} - {\rm{EFI}}{{\rm{E}}_2}} \right) +
{b_1}\left( {{\rm{N}} - {\rm{MFI}}{{\rm{E}}_1}} \right) - {b_2}\left( {{\rm{N}} -
{\rm{MFI}}{{\rm{E}}_2}} \right) = \vec 0,\\
- {c_1}\left( {{\rm{N}} - {\rm{EFI}}{{\rm{E}}_1}} \right) + {c_2}\left(
{{\rm{N}} - {\rm{EFI}}{{\rm{E}}_2}} \right) + {d_1}{\eta _1}\left( {{\rm{T}} -
{\rm{MFI}}{{\rm{E}}_1}} \right) + {d_2}{\eta _2}\left( {{\rm{T}} -
{\rm{MFI}}{{\rm{E}}_2}} \right) = \vec 0.
\end{array}
\label{eq:1}
\end{equation}
In the preceding equations, we employ the same sign conventions as in [23].
${\eta _i} = \sqrt {{\mu _i}/{\varepsilon _i}} $ is the intrinsic impedance in
region ${R_i}$ for $i = 1,2$. ${R_1}$ and ${R_2}$ are the outer and inner regions
of the scatterer, respectively. Different values can be assigned to the complex
scalar parameters ${a_i},{\rm{ }}{b_i},{\rm{ }}{c_i},{\rm{ }}{d_i}$ for $i =
1,{\rm{ }}2$ in order to obtain valid stable formulations. The expressions for
all the identities involved in Eq. (1) are the following:
\begin{equation}
\begin{array}{c}
{\rm{T}} - {\rm{EFI}}{{\rm{E}}_i}{\rm{ :~~
}}\vec 0 = {\left. {{\bf{E}}_i^{inc}({\bf{r}})} \right|_{\tan }} - {L_i}{\left.
{{\bf{J}}({\bf{r}})} \right|_{\tan }} + {K_i}{\left. {{\bf{M}}({\bf{r}})}
\right|_{\tan }} - {( - 1)^i}\frac{1}{2}{\bf{M}}({\bf{r}}) \times {\bf{\hat
n}}({\bf{r}}),\\
{\rm{T}} - {\rm{MFI}}{{\rm{E}}_i}{\rm{ :~~
}}\vec 0 = {\left. {{\bf{H}}_i^{inc}({\bf{r}})} \right|_{\tan }} - {K_i}{\left.
{{\bf{J}}({\bf{r}})} \right|_{\tan }} - \frac{1}{{\eta _i^2}}{L_i}{\left.
{{\bf{M}}({\bf{r}})} \right|_{\tan }} + {( - 1)^i}\frac{1}{2}{\bf{J}}({\bf{r}})
\times {\bf{\hat n}}({\bf{r}}),\\
{\rm{N}} - {\rm{EFI}}{{\rm{E}}_i} = {\bf{\hat n}}({\bf{r}}) \times ({\rm{T}} -
{\rm{EFI}}{{\rm{E}}_i}),~~~~{\rm{               N}} - {\rm{MFI}}{{\rm{E}}_i} =
{\bf{\hat n}}({\bf{r}}) \times ({\rm{T}} - {\rm{MFI}}{{\rm{E}}_i}).
\end{array}
\label{eq:2}
\end{equation}
In Eqs. (2), ${\bf{J}}({\bf{r}})$ and ${\bf{M}}({\bf{r}})$ denote the, a-priori
unknown, induced equivalent surface currents (electric and magnetic currents
respectively) on the interface between ${R_1}$ and ${R_2}$. ${\bf{J}}({\bf{r}})$
and ${\bf{M}}({\bf{r}})$ are vector functions of an arbitrary surface point
${\bf{r}}$, which is defined approaching the surface from ${R_1}$. Vector
${\bf{\hat n}}({\bf{r}})$ is the unit normal to the surface, pointing towards
exterior region ${R_1}$. Vectors ${\bf{E}}_i^{inc}({\bf{r}})$ and
${\bf{H}}_i^{inc}({\bf{r}})$ respectively represent the incident electric and
magnetic fields at surface point ${\bf{r}}$. ${L_i}$ and ${K_i}$ are used to
denote integro-differential operators defined as
\begin{equation}
\begin{array}{c}
{L_i}{\bf{X}}({\bf{r}}) = \int_S {\left[ {j\omega {\mu _i}{\bf{X}}({\bf{r'}}) +
\frac{j}{{\omega {\varepsilon _i}}}\nabla \left( {\nabla ' \cdot
{\bf{X}}({\bf{r'}})} \right)} \right]{G_i}({\bf{r}},{\bf{r'}})\,\,ds'} ,\\
{K_i}{\bf{X}}({\bf{r}}) = \dashint_S {{\bf{X}}({\bf{r'}}) \times \nabla
{G_i}({\bf{r}},{\bf{r'}})\,\;ds'} .
\end{array}
\label{eq:3}
\end{equation}
The symbol $\dashint$ is used in the definition of ${K_i}$ for indicating
that the integration is taken as a Cauchy principal value integral. The
integration surface $S$ refers to the separation interface between ${R_1}$ and
${R_2}$. The term ${G_i}({\bf{r}},{\bf{r'}})$ in (3) refers to the scalar Green's
function.

A generalization of the SIE--MoM formulation for the analysis of
\textit{multiple} plasmonic media can be looked up in [9], or a different
alternative approach can be found in [25]. The discretization of the unknown
currents into basis functions and the generation of the MoM linear system can be
consulted, for example, in \cite{9,23}. The comparative study included in the
following sections considers the five widespread formulations defined for the
parameters ${a_i},{\rm{ }}{b_i},{\rm{ }}{c_i},{\rm{ }}{d_i}$ in Table~1.

\ifdefined\aZ

\begin{table}[h]
\renewcommand{\arraystretch}{1.3}
\centering
\caption{Parameters for obtaining five well-documented surface integral equation formulations.}
\vskip0.2in
\begin{center}
\small
\begin{tabular}{|c||c|c|c|c|}
\hline
\textbf{\textit{formulation}} & $a_i$ for $i=1,2$  & $b_i$ for $i=1,2$ &  $c_i$ for $i=1,2$ &  $d_i$ for $i=1,2$\\
\hline
\hline
 \textbf{PMCHWT} & $\eta _i$ & 0 & 0 &  $1/\eta _i$ \\
\hline
 \textbf{JMCFIE} & 1 & 1 & 1 & 1 \\
\hline
 \textbf{CTF} & 1 & 0 & 0 & 1 \\
\hline
 \textbf{CNF} & 0 & 1 & 1 & 0 \\
\hline
 \textbf{MNMF} & 0 & ${\mu _{i}}/({\mu _{1}} + {\mu _{2}})$ & ${\varepsilon _{i}}/({\varepsilon _{1}} + {\varepsilon _{2}})$ & 0 \\
\hline
\end{tabular}
\end{center}
\label{tab:1}
\end{table}

\else

\begin{table}
\tbl{ Parameters for different well-documented SIE formulations described in \cite{20, 23}.}
{\begin{tabular}[l]{@{}lcccccc}\toprule
  formulation~~~~ & $a_i$ & $b_i$ & $c_i$ & $d_i$  \\
\colrule
  PMCHWT & $\eta_i$ & 0 & 0 & $1/\eta_i$ \\
  JMCFIE & 1 & 1 & 1 & 1 \\
  CTF & 1 & 0 & 0 & 1 \\
  CNF & 0 & 1 & 1 & 0 \\
  MNMF & 0 & ${\mu _i}/({\mu _1} + {\mu _2})$ & ${\varepsilon _i}/({\varepsilon _1} + {\varepsilon _2})$ & 0\\
\botrule
\end{tabular}}
\end{table}

\fi

\section{Iterative method overview}

The adequate choice of the iterative method to solve plasmonic problems
involving a high number of unknowns is fully supported by the fact that MoM-based
accelerating techniques --such as the fast multipole method (FMM) and the
multilevel fast multipole algorithm (MLFMA)-- are usually employed for problems
involving hundreds of thousands of unknowns. Unlike pure MoM, these MoM-based
accelerating techniques accept a controllable error in the solution [23]. In the
FMM and MLFMA algorithms, the direct inversion of the MoM matrix is not feasible,
even on modern parallel computers; and, for this reason, iterative solvers are
indispensable to find the initially unknown vector. Even when using a pure MoM
code, without FMM and MLFMA accelerations, an iterative solver usually provides
faster convergence than a direct solver such as LU matrix factorization. In fact,
the computational complexity of pure MoM with a direct solver is $O({N^3})$,
where $N$ is the number of unknowns. This complexity can be easily lowered to
$O({N^2})$ by simply switching from a direct to an iterative solver [26].

Krylov iterative solvers are among the most popular in computational
electromagnetics, because of their ability to deliver good rates of convergence
and to efficiently handle very large problems [16]. This kind of methods look for
the vector solution ${\bf{I}}$ of the system ${\bf{\bar ZI}} = {\bf{V}}$ in the
Krylov space ${K_n}({\bf{\bar Z}},{\bf{V}}) = span\left\{ {{\bf{V}},\;{\bf{\bar
ZV}},\;{{{\bf{\bar Z}}}^2}{\bf{V}}, \ldots ,\;{{{\bf{\bar Z}}}^{n - 1}}{\bf{V}}}
\right\}$, where $n = 1,2, \ldots $ represents a number of iteration in the
solver.  ${K_n}({\bf{\bar Z}},{\bf{V}})$ is a suitable space from which one can
construct approximate solutions of the linear system of equations, since it is
closely related to ${{\bf{\bar Z}}^{ - 1}}$\cite{16, 22}.

When any programmed Krylov method is called, a tolerance value $\eta $ is
inputted to the code. In practice, the solver does not run until an exact
solution is found, but rather terminate at iteration $n$ if a certain criterion
has been satisfied for the estimated solution ${{\bf{I}}_n}$. One typical
criterion is to terminate the algorithm after the following inequality is met:
\begin{equation}
{\left\| {{\bf{\bar Z}}{{\bf{I}}_n} - {\bf{V}}}
\right\|_2}/{\left\| {\bf{V}} \right\|_2} \le \eta.
\end{equation}
The term ${\left\| {{\bf{\bar Z}}{{\bf{I}}_n} - {\bf{V}}} \right\|_2}/{\left\|
{\bf{V}} \right\|_2}$ is called relative residual. Typical values for the
tolerance are $\eta  = {10^{ - 6}}$ for double precision (64 bits) entries in
${\bf{\bar Z}}$ and ${\bf{V}}$, and $\eta  = {10^{ - 3}}$ for single precision
(32 bits) entries. However, for MoM-based simulations involving tens of millions
of unknowns, the tolerance must be raised to values as high as about $\eta  =
{10^{ - 2}}$ [27]. This increase in $\eta $ is needed because the condition
number of the MoM system typically worsens as the number of unknowns lifts and
the number of iterations to achieve $\eta $ would be impracticable to reach.

The Krylov method GMRES (generalized minimum residual method), developed by Y.
Saad and H. Schultz in 1986 [21], is known to be the optimal iterative solver in
the sense that it minimizes the number of iterations required to converge
satisfying (4) [16]. Nevertheless, the optimality of GMRES comes at a price. The
memory cost of applying the method increases with the iterations, and it may
sometimes become prohibitive for solving certain problems. As an attempt to limit
this cost, there exist restarted versions of GMRES in which, after a given number
\textit{R} of iterations (\textit{R} is known as the restart parameter), the
approximate solution for the next steps is computed form the previously generated
Krylov subspace. Then this existent Krylov subspace is completely erased from the
memory, and a new space is constructed from the latest residual. In a restarted
GMRES, each ordinary iteration is called internal iteration, whereas a set of
\textit{R} (\textit{R} denotes again the restart value) internal iterations is
called external iteration. Summarizing the storage costs, the number of
additional floats or doubles --over the baseline memory requirements in
MoM/FMM/MLFMA-- that GMRES requires in memory can be easily estimated from the
following analytical results [28]:
\begin{equation}
GMES\_RAM\_overuse(M,N){\rm{  }} = {\rm{   }}2MN + {M^2} - 7M~~{\textrm{
 floats or doubles}},
\end{equation}
where \textit{N} is the number of complex unknowns and $M$ is the number of
iterations required to achieve (4) in the unrestarted ordinary GMRES. For a
restarted GMRES, $M$ is computed from the restart parameter \textit{R} as $M =
\min (P,R)$, with $P$ the total number of internal iterations.

In addition to the restarted versions of GMRES, other non-optimal iterative
solvers attempt to preserve the favorable convergence properties of GMRES while
introducing a negligible overuse in RAM usage. For this paper, we consider
complex versions of iterative Krylov solvers based on different versions of the
so-called Lanczos biorthogonalization algorithm [22]: TFQMR (transpose free
quasi-minimal residual method), CGS (conjugate gradients squared method) and
BICGSTAB (bi-conjugate gradients stabilized method).

\section{Comparative study of numerical results}

The analysis of the accuracy and the iterative performance of the five SIE
formulations described in Section 2 is carried out in this section for
representative problems involving plasmonic materials. Error and iterative
performance results were obtained for the Krylov solvers described in Section 3.
The error analyses are summarized in Subsection 4.1. Then, in Subsection 4.2, we
present iterative performance analyses that not only include spheres as
scatterers, but also real models of plasmonic nanoantennas. The GMRES solver with
a fixed restart was the only solver considered in [20], whereas our results for
plasmonic problems cover the following important cases: three additional solvers
(TFQMR, CGS and BICGSTAB), the ordinary GMRES without restarts, and the restarted
GMRES including a set of different restart parameters not considered in the
aforesaid reference. Furthermore, the results in [20] cover up to 36,000
unknowns. In contrast, we have extended the comparative results up to around
100,000 unknowns.

Some quantitative results presented in this paper regarding the iterative
performance and the accuracy differ from their counterparts in [20] --even though
there is a general agreement among those main qualitative results that are common
to both papers-- due to two major reasons: i) we employed a Gaussian quadrature
rule, described in [29], consisting of 7 points per triangle for numerical
integration, whereas in [20] a 3-point rule was used; ii) unlike the simulations
in [20], we employed a diagonal preconditioner described in [29] together with
the preconditioning technique in [30]. A diagonal preconditioner and the
technique in [30] are straightforward to implement in any existing MoM code
without adding significant computational complexity. Moreover, in 10.2.4 of [29]
is stated that experimental results show that the 7-point quadrature rule is
preferable to the 3-point rule, as it provides superior precision for MoM
problems, which strengthens our choice for this integration technique.

This paper also describes the numerical precision and meshes employed for the
discretization, not clearly stated in some of the aforementioned literature, in
order to allow a full reproducibility of our results by the scientific community.
Double-precision floating-point C calculations were used for all the results in
this paper, and we employed the so-called ``frontal'' mesh type in the free
software Gmsh [31]. The ``frontal'' type was selected because it provides
triangles with good aspect ratios, namely, triangles which do not have small
internal angles. This triangle feature is required in MoM to obtain accurate
results, as explained in 8.7.4.1 of [29].

For all the studied iterative solvers, the maximum number of external iterations
was unlimited --with the exception of the unrestarted GMRES, where the number of
iterations is limited to the number of unknowns--, and the relative residue
tolerance for stopping each method was set to 10$^{-6}$, in accordance with the
traditionally used tolerance value which can be found, for instance in \cite{20,30}.

The main graphical representations of the results in this paper were obtained
with a pure MoM implementation, as in [20]. However, some figures in this paper
include MLFMA results whenever a ``MLFMA vs. pure MoM'' comparison is relevant.
The MLFMA implementation used for the present work was configured with the same
MLFMA parameters described in [32]. Our computational implementation of the
MoM/MLFMA-SIE method consists of a regular C code involving double-precision
floating-point calculations. For the accurate evaluation of the singular
integrals, we have used all the proper analytical extraction techniques explained
in [29]. In order to model the MoM currents, we have employed the well-known
Rao-Wilton-Glisson (RWG) basis functions [29]. The Galerkin's method was assumed
for this work, meaning that the MoM testing functions are the same as the basis
functions.

All our simulations were carried out on a computer with two
Intel\textregistered{} Xeon\textregistered{} Processors E5-2690v2 running the
64-bit operating system Windows 8.1 Professional. The code was not parallelized
by hand, but automatically parallelized using the source-to-source compiler
Parallware [33] to be run with 20 threads executed on half of the cores. The
aforementioned model of the processors and the number of threads have no
influence on any comparative result, but only determine the absolute quantitative
runtime outcomes in Subsection 4.2. Regarding the automatic parallelization,
Parallware is a parallelizing tool that automatically extracts the parallelism
implicit in the source code of a sequential simulation program written in the
regular C programming language. In addition, Parallware automatically generates an optimized
parallel-equivalent program written in C and annotated with OpenMP [34] compiler
pragmas.

\subsection{Assessment of numerical accuracy}

A comparative analysis of the accuracy was carried out employing the five
selected MoM-based SIE formulations and the four solvers. The same sphere with
radius ${\lambda _0}/2$ employed in 3.3 of [20] was used in this section as a
representative example for evaluating the accuracy of each considered iterative
solver. This sphere is illuminated by a plane wave, with incidence direction $
{\bf{\hat z}}$ and polarization ${\bf{\hat x}}$, at the operating optical
wavelength ${{\lambda _0}=548.6{\rm{~nm}}{\rm{.}}}$ The following plasmonic
materials were chosen for the sphere composition: gold (${\varepsilon _r} =  -
5.8 - j2.1$, at the simulation frequency), silver (${\varepsilon _r} =  - 12.8 -
j0.4$) and aluminum (${\varepsilon _r} =  - 35.2 - j9.82$). The values of the
complex relative dielectric permittivity for each material have been extracted
from [20].

The normalized root mean square (RMS) error with respect to the Mie's series
results was calculated using the following expression:
\begin{equation}
{e_{rms}} = \left( {\sqrt {\sum {{{(E_{ref}^{} - {E_{simul}})}^2}/N} } }
\right)/\max \left\{ {E_{ref}^{}} \right\},
\label{eq:4}
\end{equation}
where ${E_{simul}}$ is the magnitude of the scattered electric far field
obtained in the simulation for $N$ different observation values. ${E_{ref}}$ is
the reference field provided by the Mie's series. For far-field patterns, the
error was calculated using $N = 360$ equispaced angular values for the variable
$\phi $ on plane XZ. The near-field patterns comprise a mesh with resolution $200
\times 200$ points. This mesh was created on plane XZ over a centered square of
side length $4r$, with $r$ the sphere radius.

We found no significant differences in the comparative error variation among the
three considered plasmonic materials. We did not either encounter any effect of
the chosen iterative solver on the normalized RMS error, as all the four solvers
yielded identical error levels. {
These results in this section complete the important findings published in \cite{20}, extending the error analysis to multiple iterative solvers.
}
Fig. 1 shows the error versus the number of
unknowns for each SIE formulation when the sphere is made of gold. The variation
on the number of unknowns was achieved by varying the maximum side length ${\ell
_{discr}}$ of the triangles in the geometry mesh, according to the rule ${{\ell
_{discr}} = \frac{{{\lambda _0}}}{{10n}},{\rm{ }}n = 1,2, \ldots {\rm{ }}7.}$

\begin{figure}[h]

\centerline{\includegraphics[width=0.7\columnwidth,draft=false]{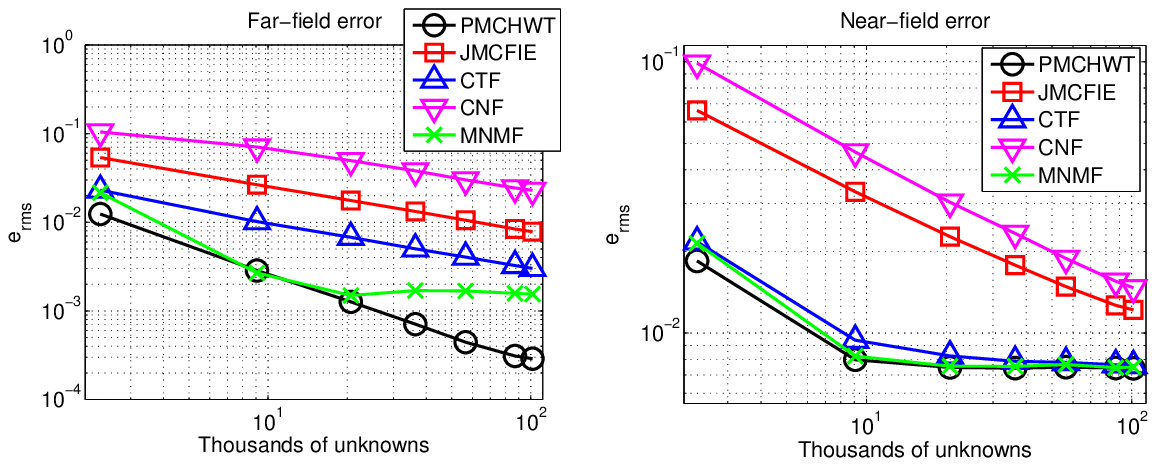}}

\caption{Comparative study of the RMS error for a golden sphere with radius $r =
{\lambda _0}/2$: (left) far field; (right) near field. Error  values  were  found
 to  be identical  among  the  four  considered iterative  solvers.} 

\end{figure}

In terms of accuracy, the PMCHWT has clearly shown to be the most reliable
formulation in plasmonics. This statement is in full agreement with the results
in [20]. The MNMF formulation has an error level comparable to the PMCHWT in most
cases; however its far-field error level becomes unstable and starts to grow for
high numbers of unknowns. This anomalous behavior in MNMF might be ascribed to
the singularity extraction techniques involved in the MNMF implementation [29].
The CTF formulation exhibits a near-field error similar to the PMCHWT, but in the
far field its error level is notably worse. Given that both PMCHWT and CTF
combine tangential equations only, the accuracy provided by the PMCHWT highlights
the actual importance of the combination scalar parameters on the obtained level
of accuracy.

As a graphical reference, Fig. 2 shows far and near-field reference values
obtained from the Mie's series for the sphere with radius ${\lambda _0}/2$.
PMCHWT far-field values simulated with two very different numbers of unknowns
(mesh sizes) are represented in Fig. 2 in order to allow for visual assessment of
the real importance of taking into account the error variations shown in Fig. 1.

\begin{figure}[h]
\centerline{\includegraphics[width=0.7\columnwidth,draft=false]{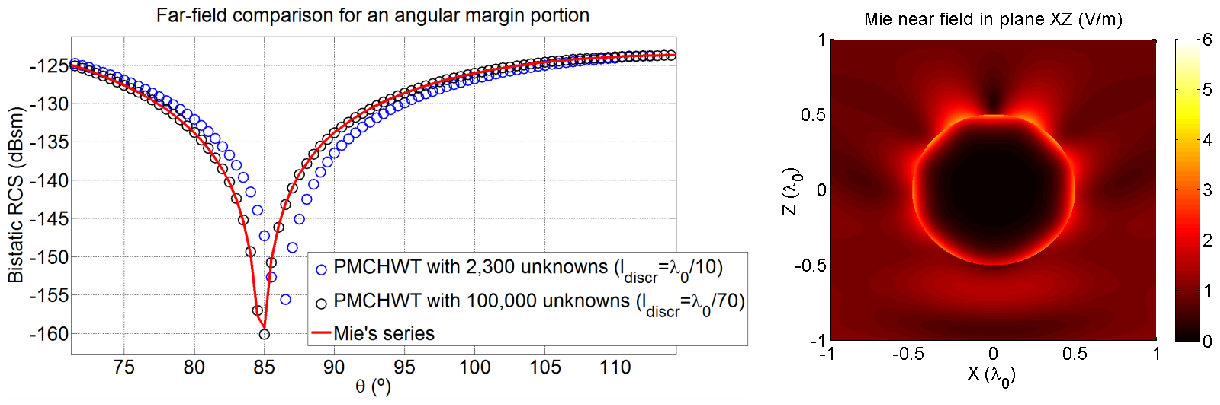}}
\caption{Reference results from Mie's series: (left) far field; (right) near
field. The far-field results include a comparison with PMCHWT predictions for two
different representative mesh sizes.} 
\end{figure}

\subsection{Assessment of iterative performance}

With the purpose of assessing the iterative performance of the five studied SIE
formulations, we initially simulated the ${\lambda _0}/2$ radius gold sphere,
using MoM solved iteratively. We employed four well-known iterative solvers for
linear systems of equations [22]: TFQMR, CGS, BICGSTAB and GMRES --an unrestarted
GMRES version, plus three more GMRES versions with restart parameters 30, 60 and
90--. In Fig. 3, we analyze the influence of the GMRES restart parameter on the
memory cost. This analysis is used to justify our elections for the mentioned
restart values. Fig. 3 shows on the left the memory usage required by pure MoM
and also by MLFMA, as a function of the number of unknowns. These curves on the
left are independent of the particular chosen solver. Then, on the right of Fig.
3, the RAM overuse introduced by GMRES is represented for the restart values 30,
60 and 90, and for the unrestarted version.

As inferred from Fig. 3, the percentage RAM overuse may be negligible when using
pure MoM together with unrestarted GMRES, but is noticeably worse in the MLFMA
case. In other words, even if the available RAM allows a MLFMA simulation, the
election of the unrestarted GMRES for the solver may make this simulation
unfeasible. The maximum GMES restart considered in the present work,
\textit{R}=90, was selected to bound the total RAM consumption increase in MLFMA
to around 1\% of the total RAM usage in every non-GMRES Krylov iterative solver.

\begin{figure}[h]
\centerline{\includegraphics[width=0.7\columnwidth,draft=false]{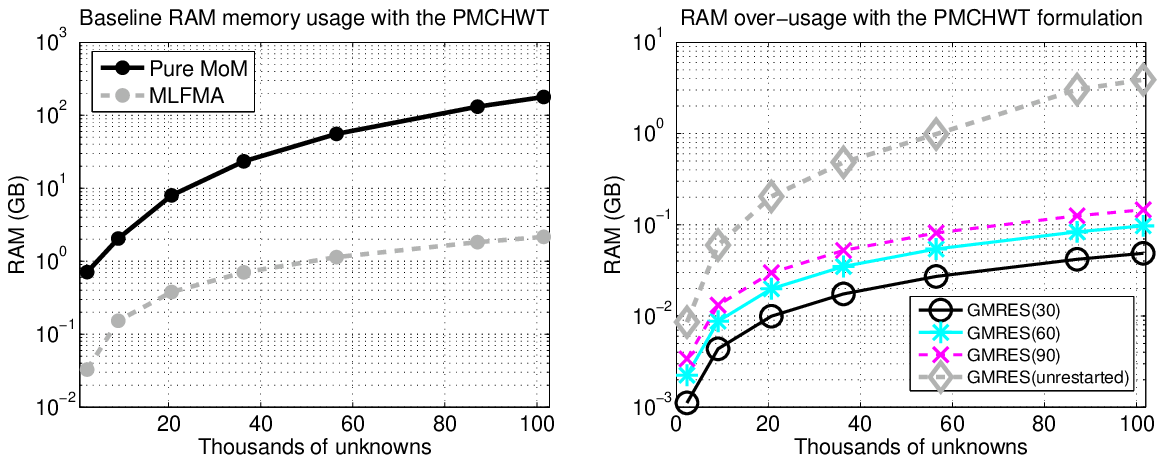}}
\caption{RAM memory usage with GMRES: (left) baseline memory usage in pure MoM
and in MLFMA, without considering iterative solver requirements; (right) RAM
consumption overuse for different GMRES versions with varying restarts.} 
\end{figure}

Measurements of the iterative performance of the formulations and solvers are
shown on the left of Fig. 4, where the runtime for iteratively solving the MoM
system has been taken from the fastest iterative solver for each formulation. For
this particular comparison, unrestarted GMRES was not considered, in order to
make a fair comparison among formulations. Indeed, for every formulation, GMRES
is optimal; and, as a consequence, it outperforms all other Krylov methods in
terms of iterative performance. However, as show above, unrestarted GMRES should
be only used when memory requirements are not a concern.

Fig. 4 also includes on the right, for comparison purposes, the time required to
fill in the full MoM matrix before the iterative solver begins. The time for
filling the MoM matrix is slightly smaller with the PMCHWT and the CTF because,
in these formulations, it is possible to avoid the integration of the Green's
function gradient which appears in operator ${L_i}$ in Eq. (3). This
simplification on the integral requires employing the two-dimensional version of
the divergence theorem, which cannot be invoked for the formulations that involve
normal equations, due to the vector product by ${\bf{\hat n}}({\bf{r}})$.

As shown in Fig. 4, the restarted GMRES(90) solver is the best choice for all
the formulations, except for the tangential formulations PMCHWT and CTF, where
TFQMR provides the best iterative behavior. Another important comment on Fig. 4
is that the MNMF formulation provides the best performance when solved
iteratively.

\begin{figure}[h]
\centerline{\includegraphics[width=0.7\columnwidth,draft=false]{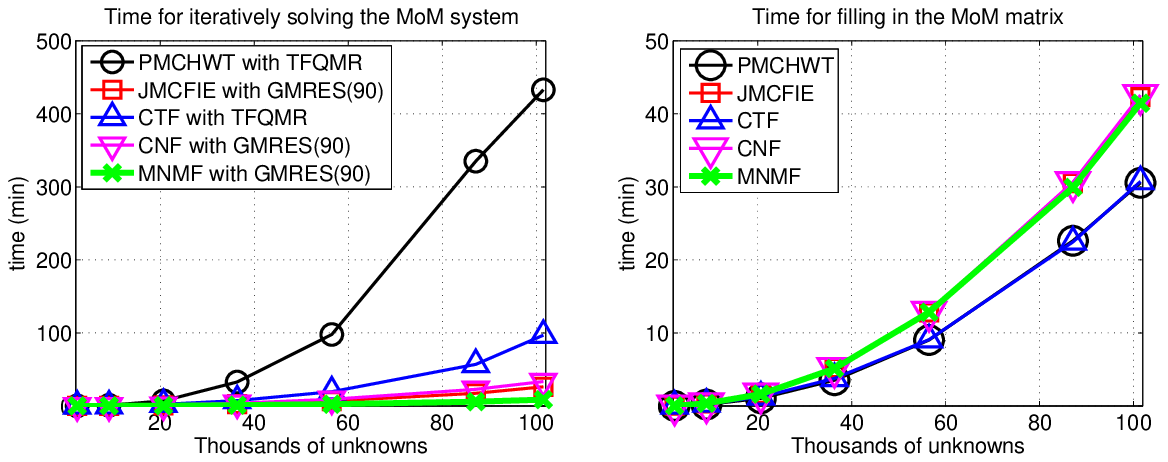}}
\caption{Runtimes for simulating the ${\lambda _0}/2$ radius gold sphere: (left)
runtime for iteratively solving the MoM system when the iterative method with the
fastest convergence is chosen for each SIE formulation; (right) total runtime for
filling in the MoM matrix.} 
\end{figure}

Since, as discussed above in Subsection 4.1, the PMCHWT provides the best
accuracy when dealing with plasmonic problems, we chose to represent in Fig. 5
the iterative performance of this formulation for each solver, including GMRES
without restart. {
Even though important previous works such as \cite{15} indicate that the PMCHWT is unfit to be solved by an iterative method, the scenario considered in
this paper is different and not directly comparable to \cite{15}. The recently introduced preconditioning scheme in \cite{30}, not yet available at the time of the report in \cite{15}, was considered for the simulations in this work, as justified above.
}

As can be seen in the graphical results, CGS and TFQMR are the
best choices for the PMCHWT, if unrestarted GMRES is excluded owing to memory
requirements. Fig. 5 also shows that the number of matrix-vector multiplications
(MVMs) required by each iterative solver is proportional to the runtime. It is
worth stating that all the solvers require two MVMs in each iteration, except for
GMRES (all versions), which requires only one MVM per internal iteration. The
number of MVMs is thus a machine-independent parameter that can be easily used to
compare the total processing time associated with each one of the formulations
and solvers.

Let us note that for the formulations other than the PMCHWT all the solvers
display a similar iterative behavior, as seen in Fig. 6.

\begin{figure}[h]
\centerline{\includegraphics[width=0.7\columnwidth,draft=false]{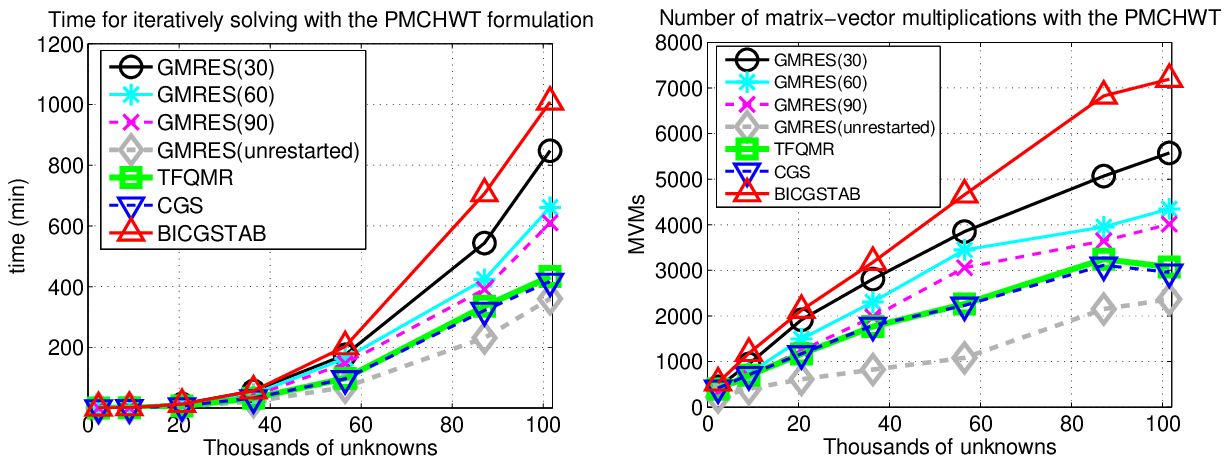}}
\caption{Iterative performance for the PMCHWT formulation in the considered Mie
scattering problem: (left) time for solving the MoM system; (right) number of
matrix-vector multiplications.} 
\end{figure}
\begin{figure}[h]
\centerline{\includegraphics[width=0.7\columnwidth,draft=false]{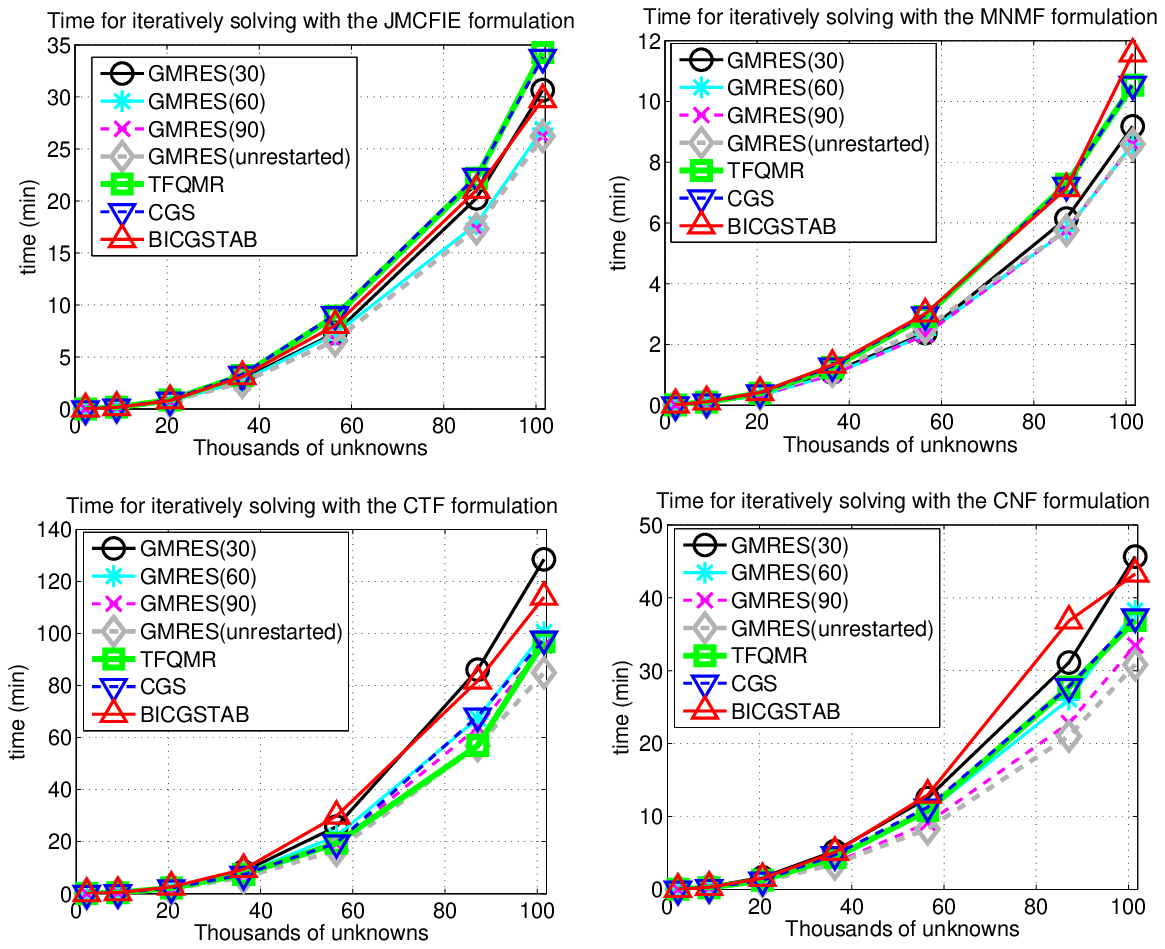}}
\caption{Iterative performance for the JMCFIE, MNMF, CTF and CNF formulations.} 
\end{figure}

Fig. 7 shows the relative residual as a function of the number of MVMs
(matrix-vector multiplications) for the PMCHWT formulation for 57,000 unknowns
--other mesh sizes exhibit an analogous behavior--. As expected, GMRES without
restart exhibits the fastest residual decay.
{
Regarding data from the CGS and BICGSTAB solvers in Fig. 7, it can be noted that the time-dependent residual variances are very high. A similar erratic residual evolution can be also observed in the iterative solvers included in commercial codes like FEKO; nonetheless, the important fact is that the time-averaged minimum residual is reduced. As a matter of fact, it must be clear from Fig. 7 that, for instance, both TFQMR (non-erratic) and CGS (erratic in this scenario) exhibit a similar time-averaged minimum residual evolution and they converge at a similar rate.
}

Returning to Fig. 7, it is clear that for the cases where memory-expensive
unrestarted GMRES is not an option, TFQMR and CGS provide a better convergence
than GMRES(90) in the PMCHWT, for the residue tolerance of 10$^{-6}$; however,
this will not be the case if a higher residual error is defined to achieve
convergence. This observation regarding the PMCHWT formulation is
particularly important because, when programming a single SIE formulation in a MoM
code, the PMCHWT is generally
chosen as the preferred formulation due to the following reasons: 1) it is the
easiest to implement, as it does not require using the normal vector directions;
and 2) it is employed in commercially available software like FEKO [35] and in
publications about simulating plasmonic bodies such as [8].
\begin{figure}[h]
\centerline{\includegraphics[width=0.7\columnwidth,draft=false]{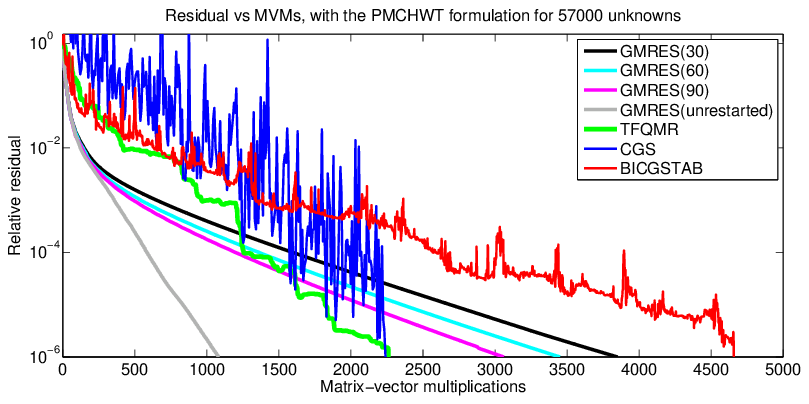}}
\caption{ Relative residue versus matrix-vector products in the PMCHWT
formulation, for each iterative solver.} 
\end{figure}

In order to extract valid general conclusions about the iterative runtimes, we
also analyzed spheres made of silver and aluminum and we tested more
elaborated geometries at different optical frequencies, such as a nanoantenna
made of aluminum.  { These general conclusions must be understood as
valid under the simulation parameters justified at the beginning of Section 4. The use of
alternative preconditioners would certainly be an important point to consider in order to obtain more general conclusions; however, these preconditioners would have required a noticeably extended explanation out of the scope of this paper. Likewise, despite the fact that higher-order basis functions are not indispensably required to perform accurate plasmonic simulations \cite{20}, we understand that they could be interesting for further research. Additionally, sharp edges and corners are not considered in the present work when extracting general conclusions. It is well known that sharp edges and corners are not typically found in real plasmonic structures chemically generated in a laboratory setup \cite{11}; however we believe that additional research should be pursued in the future regarding plasmonic edges in both experimental and theoretical realms. Finally, following a similar approach used in the literature \cite{18}, we have only represented the most relevant cases when dealing with qualitative conclusions. The results corresponding to the spheres made of silver and aluminum, as well as simulations varying the nanoantenna plasmonic composition, are not shown for the sake of simplicity, as these results have proved to be qualitatively equivalent to the shown ones.

}

 The considered
real plasmonic Yagi-Uda nanoantenna for the emission of light at ${\lambda _0} =
570~{\rm{ nm}}$, whose emitted near field pattern obtained with our code is represented in Fig.~8, has been designed in [5]. See this reference for a detailed
description of the antenna dimensions, which have been optimized for high optical
directivity at the operating frequency. The optical antenna is made of aluminum
(${\varepsilon _r} =  - 38 - j10.9$ at the simulation frequency), and it enhances
the emission of a single fluorescent chemical molecule modeled as a classical
hertzian dipole along direction ${\bf{\hat y}}$, 4~nm above the feed element. In
our particular simulation, the \textit{equivalent} dipole has length 8 nm and its
electric current is 1 nA.

\begin{figure}[h]
\centerline{\includegraphics[width=0.7\columnwidth,draft=false]{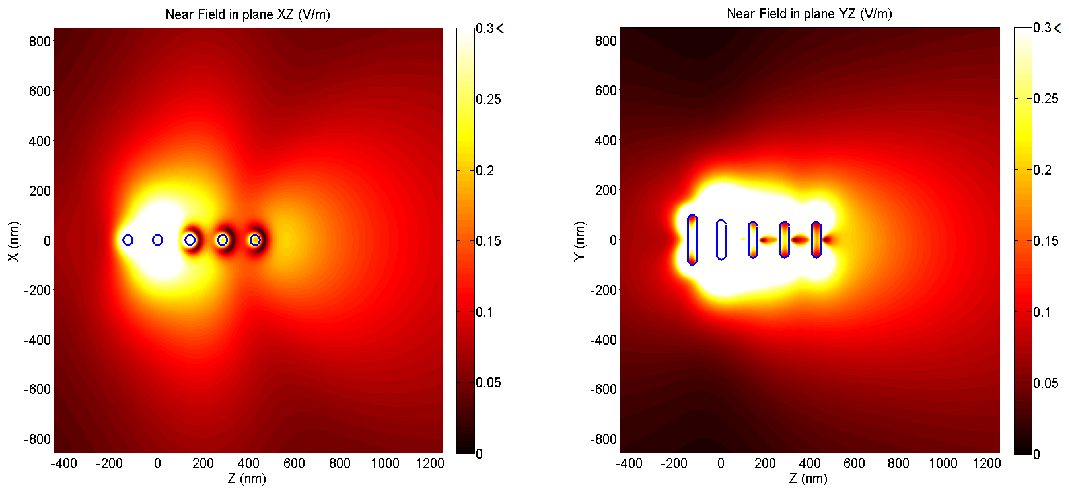}}
\caption{Near field pattern emitted by the aluminum nanoantenna designed in [5].
The represented near-field magnitudes were obtained with the PMCHWT formulation
for a mesh size ${\ell _{discr}} = {\lambda _0}/40$.} 
\end{figure}

As inferred from Fig. 9, iterative performance in the nanoantenna problem is
similar to that previously presented for the Mie scattering problem when the
sphere was simulated. The absolute runtimes versus the number of unknowns vary
when compared to the Mie scattering, but the qualitative differences among the
four iterative solvers remain very similar, which allows extracting some general
conclusions for a wide variety of tests involving plasmonic media. These
conclusions are summarized in the next section.

\begin{figure}[h]
\centerline{\includegraphics[width=0.7\columnwidth,draft=false]{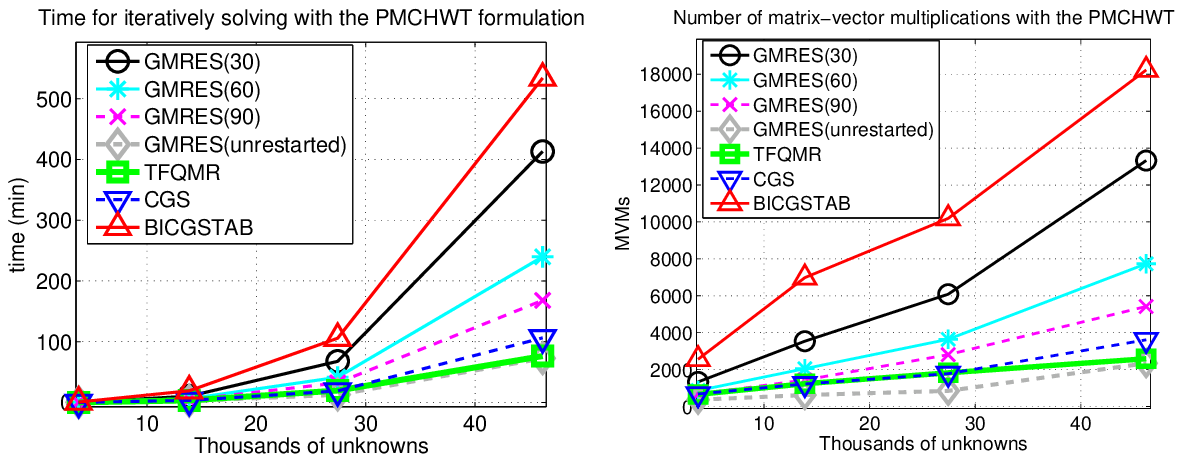}}
\caption{Iterative performance for the PMCHWT formulation when the aluminum
nanoantenna in [5] is simulated: (left) time for solving the MoM system; (right)
number of matrix-vector multiplications.} 
\end{figure}

\section{Conclusions}

Four well-documented iterative solvers have been applied to five widespread
MoM-SIE approaches in the analysis of nanostructures made of plasmonic materials.
In order to increase the reproducibility of the results in this paper, we have
provided in-depth details about all the parameters involved in our code
implementation (numerical integration rule, mesh type, computational precision,
etc.), and we have explained the reasons and convenience for the election of such parameters.
These thorough explanations on code implementation are usually not described in
detail, and they affect the reproducibility of results in previous papers about
iterative solvers in computational electromagnetics. Our observations are
summarized as follows:

\begin{enumerate}

\item In the case where memory requirements are not an impediment according to
Eq.~(5), then the best election for all the formulations is always GMRES without
restart.

\item Whenever accuracy is required in the results, then the best choice is to use the
PMCHWT formulation, because this formulation has the smallest error in all the
analyzed plasmonic problems; however, the PMCHWT has a poor iterative
performance.

\item If the PMCHWT is chosen and unrestarted GMRES cannot be applied due to memory
requirements, then the best iterative solvers are TFQMR and CGS, because they may
provide a much better performance, for the standard double-precision residue
tolerance of 10$^{-6}$, than the rest of the analyzed solvers. TFQMR and CGS are
also the best choices for the CTF formulation (although in this case the effect
is less noticeable).

\item The fastest convergence in plasmonics belongs to the MNMF and JMCFIE formulations. For these
formulations, importantly, however, the choice of the iterative solver has a small impact on
the total runtime.

\item For a given fixed residual tolerance and iterative solver, the choice of the SIE formulation has a direct impact on
the simulation error and also an impact on the runtime. Nonetheless, for a fixed residual tolerance and SIE formulation,
the choice of the iterative solver has a negligible impact on the
simulation error but, in general, a relevant impact on the total runtime.

\end{enumerate}

{
In this paper, we did not only confirm conclusions in items 1 and 2 above --also approached in \cite{16} and \cite{20} respectively-- but we have also extended our analysis to extract the remaining conclusions on the list above. These novel conclusions, together with the quantified results in the presented figures, are relevant contributions from this work to the know-how literature of computational electromagnetics in the context of plasmonic problems.

}

\section*{Acknowledgement}

The authors especially thank the company \textit{Appentra Solutions,} developers
of the automatic parallelizing source-to-source compiler \textit{Parallware}
(www.appentra.com), for assisting us in the analysis and parallelization of some
parts of our C codes.

\section*{Funding}

This work is partially supported by the Spanish National Research and
Development Program under project TEC2011-28683-C02-02, by the Spanish Government
under project TACTICA, by the European Regional Development Fund (ERDF), and by
the Galician Regional Government under agreement for funding AtlantTIC (Atlantic
Research Center for Information and Communication Technologies).

\end{document}